\documentclass{elsart}

\usepackage{graphicx}
\usepackage[latin1]{inputenc}
\usepackage{amsmath}
\usepackage{amsfonts}
\usepackage{amsbsy}
\usepackage{amssymb}

\newcommand{\lap} {\Delta}

\renewcommand{\div} {\nabla\cdot}
\renewcommand{\vec}[1] {{\textbf{\textit{#1}}}}

\renewcommand{\u} {\textbf{\textit{u}}}
\newcommand{\x} {\textbf{\textit{x}}}
\renewcommand{\k} {\textbf{\textit{k}}}
\renewcommand{\e} {\textbf{\textit{e}}}

\begin{document}

\begin{frontmatter}

\title{Impact of the floating-point precision and interpolation scheme
on the results of DNS of turbulence by pseudo-spectral codes }

\author{Holger Homann, J\"urgen Dreher and Rainer Grauer}

\address{Institut für Theoretische Physik I,
         Ruhr-Universität-Bochum,
         44780 Bochum, Germany}

\begin{abstract}
  In this paper we investigate the impact of the floating-point
  precision and interpolation scheme on the results of direct
  numerical simulations (DNS) of turbulence by pseudo-spectral codes.
  Three different types of floating-point precision configurations
  show no differences in the statistical results. This implies that
  single precision computations allow for increased Reynolds numbers
  due to the reduced amount of memory needed.  The interpolation
  scheme for obtaining velocity values at particle positions has a
  noticeable impact on the Lagrangian acceleration statistics. A
  tri-cubic scheme results in a slightly broader acceleration
  probability density function than a tri-linear scheme. Furthermore
  the scaling behavior obtained by the cubic interpolation scheme
  exhibits a tendency towards a slightly increased degree of
  intermittency compared to the linear one.
\end{abstract}

\begin{keyword}
Direct numerical simulations \sep Spectral methods \sep Interpolation
methods \sep Turbulent flows \sep High-Reynolds-number turbulence \sep
Isotropic turbulence; homogeneous turbulence
 
\PACS 
47.27.ek
%47.27.ek Direct numerical simulations
\sep
47.27.er
% 47.27.er Spectral methods
\sep
02.60.Ed
%02.60.Ed Interpolation methods
\sep
47.27 -i
% 47.27.-i Turbulent flows
\sep
47.27.Jv
% High-Reynolds-number turbulence
\sep
47.27.Gs
%47.27.Gs Isotropic turbulence; homogeneous turbulence
\end{keyword}
\end{frontmatter}

% main text

\section{Introduction}
Due to the increasing speed and memory of computers, numerical
simulations of the Navier-Stokes equations have become a valuable tool
for studying turbulence. To investigate the intrinsic properties of
turbulence such as the energy cascade and intermittency one often uses
periodic boundary conditions in order to keep the influence of the
geometry on the flow structure as small as possible. For this kind of
turbulence simulations pseudo-spectral codes are widely used
\cite{gottlieb-orszag:1977,canuto-etal:1988,vincent-meneguzzi:1991,she-chen-etal:1993,mueller-biskamp:1999,DBLP:conf/sc/YokokawaIUIK02,chevillard:2005,yeung-borgas:2004,biferale:2004b,bec-etal:2006}.
These treat the Navier-Stokes equations in Fourier space and compute
the convolutions in real space. A fast Fourier transformation is used
to switch between the two spaces.

A fundamental feature of turbulence is the inertial range of scales.
Within this range energy solely cascades by the inertial interaction
of eddies, which causes the turbulence to be
scale-independent. Attached to this range at small scales is the
dissipation range where dissipative effects are important and damp the
turbulent motion. The size of the inertial range is directly connected
to the Reynolds number $Re = VL/ \nu$, where $V$ and $L$ denote the
velocity and size of the large scale motion and $\nu$ the kinematic
viscosity of the fluid. Many theories of turbulence deal with an
infinite Reynolds number \cite{kolmogorov:1991} which means that they
focus on the properties of the inertial range. Unfortunately the
Reynolds number $Re$ is connected to the degrees of freedom $N$ of the
turbulent flow according to $Re \sim N^{9/4}$. Therefore the memory
and speed of computer limits the achievable Reynolds number and size
of the inertial range in direct numerical simulations.

Nearly all numerical simulations are performed with double
floating-point precision.  Recently, the largest numerical simulation
worldwide \cite{DBLP:conf/sc/YokokawaIUIK02} performed on the Earth
Simulator used single precision data for the velocity field and double
precision data for the calculation of the convolution sums in order to
reduce the amount of memory needed. For this reason it was possible to
set up a simulation of $4096^3$ grid points.

In this paper we examine the impact of the floating-point precision on
the numerical results. Therefore we performed simulations using three
different configurations of floating-point precision (see
Table~\ref{precisionTable}).  The first configuration, which is the
most common approach, computes all fields with double precision
(RUN1). The second corresponds to the configuration on the Earth
Simulator which uses single precision for the velocity fields and
double precision for the convolutions (RUN2). The third uses single
precision for all fields (RUN3), which halves the needed amount of
memory compared to RUN1 and therefore allows for an increased Reynolds
number.
\begin{table}[t]
  \centering
  \begin{tabular}{c|ccc}
         & fields           & convolutions     & interpolation \\ \hline
    RUN1 & double precision & double precision & tri-cubic \\ 
    RUN2 & single precision & double precision & tri-cubic \\ 
    RUN3 & single precision & single precision & tri-cubic \\ 
    RUN4 & double precision & double precision & tri-linear
  \end{tabular}
  \caption{\label{precisionTable} List of floating-point precision
    configurations and interpolation schemes for RUN1 -- RUN4.}
\end{table}

Interesting quantities in turbulence are obtained from statistical
averages in space such as the energy spectrum and Eulerian structure
functions or in time such as the acceleration statistics of passive
tracers and Lagrangian structure functions. Especially for the
Lagrangian statistics a lot of effort has been invested into numerical
simulations
\cite{chevillard:2005,yeung-borgas:2004,biferale:2004b,bec-etal:2006}
and experiments \cite{mordant:2004,porta-bodenschatz-etal:2001} of
tracers and heavy particles and in turbulent flows. The integration of
the tracers (and particles) requires the interpolation of the velocity
field at the tracer positions. In order to obtain reliable Lagrangian
statistics it is necessary to achieve accurate trajectories from the
simulations.  The crucial point is the interpolation scheme
used. There are three different schemes applied in the literature. The
first is based on cubic splines
\cite{chevillard:2005,yeung-borgas:2004}, the second is a tri-cubic
scheme \cite{busse-homann-mueller-grauer:2006} and the third a
tri-linear scheme \cite{biferale:2004b}. In this paper we investigate
the impact of the interpolation scheme on the Lagrangian statistical
results. The scheme using splines has the drawback that it is
difficult to parallelize for distributed memory computer due to its
lack of locality. Most of the current massive parallel computers are
of this type of memory so we will not consider interpolation schemes
using splines. For comparison we performed a simulation with a
tri-cubic (RUN1 -- RUN3) and a tri-linear (RUN4) interpolation scheme.

\section{The Numerics}
The simulations are performed by numerically solving the
incompressible Navier-Stokes equations,
\begin{align}
  \label{impuls}
  \partial_t \u + (\u \cdot\nabla)\u &= -\nabla p +\nu \lap \u, \\
  \label{konti}
  \div \u &= 0,
\end{align}
in a periodic cube with $256^3$ collocation points with a
pseudo-spectral method using spherical mode truncation to reduce
aliasing effects \cite{vincent-meneguzzi:1991}. The non-linear term in
(\ref{impuls}) would be a convolution in Fourier space and therefore
is treated in real space. The dissipation term can be computed exactly
in Fourier space. The equation of continuity (\ref{konti}) determines
the pressure $p$ and is satisfied by first neglecting the pressure
term and afterwards projecting the velocity on its solenoidal part.

We parallelize the computations via slab geometry. This means every
process holds the data of a sub-slice of the whole cube. The parameters
of the simulations RUN1 -- RUN4 are listed in Table~\ref{table}.
\begin{table}[t]
  \centering
  \begin{tabular}{ccccccccccccc}
    $Re$&$u_0$ & $\epsilon$ &  $\nu$          &  dx    &          $\eta$      &$\tau_\eta$&$L$&$T_L$& $N^3$ & $N_p$\\\hline
%    1575&$0.15$&$1.5\cdot 10^{-3}$&$2 \cdot 10^{-4}$&$2.45 \cdot 10^{-2}$&  $8.6\cdot 10^{-3}$  &  0.37  &2.1&  14    &$256^3$& $1 \cdot 10^5$
    1575&$0.15$&$0.0015$    &$0.0002$&$0.0245$&  $0.0086$  &  0.37     &2.1&  14 &$256^3$& $1 \cdot 10^5$
  \end{tabular}
 \caption{\label{table}
    Parameters of the numerical simulations.
    $Re$: Reynolds number $VL/ \nu$,
    $u_0 = \sqrt{2/3 E_k}$, $E_k$: kinetic energy,
    $\epsilon$: mean energy dissipation rate,
    $\nu$: kinematic viscosity, 
    $\eta$: dissipation length scale $(\nu^3/\epsilon)^{1/4}$,
    $\tau_\eta$: Kolmogorov time scale $(\nu/\epsilon)^{1/2}$,
    $L = (2/3 E)^{3/2}/\epsilon$: integral scale,
    $T_L = L/u_0$: large-eddy turnover time,
    $N^3$: number of collocation points,
    $N_p$: number of particles}
\end{table}
For the inter-process communication we use the Message Passing
Interface (MPI). The FFTs are performed by the MPI-parallel C-library
FFTW (version 2.1.5) \cite{fftw}. The time integration of the velocity
field is done by a strongly stable Runge-Kutta third order scheme
\cite{shu-osher:1988}.

In order to advance the tracers the velocity field has to be
interpolated at the tracer positions. The velocity field is given on a
equally spaced grid and evolves according to the Navier-Stokes
equations. The tracers have to be integrated at run-time due to the
limited amount of hard-disk space. This is done using the same
Runge-Kutta third order scheme as for the integration of the velocity
field. To interpolate the velocity field at the positions of the
individual particles, we implemented a tri-cubic and a tri-linear
scheme. The tri-cubic interpolation relies on prescribing the values of
the function, its first, mixed second and mixed third derivative at
the eight corners of the particle surrounding cubic grid cell (see
\cite{press-teucholsky-etal:1992} for the 2D version). For calculating
the derivatives we use a first order centered scheme. The tri-linear
interpolation just needs the values of the function at the corners.
Both schemes parallelize very efficiently.

The turbulence is driven by keeping constant the Fourier modes with $1
\le \arrowvert \k \arrowvert \le 2$. The starting point of the
different runs is the same snapshot of a statistically stationary
flow. Subsequently we injected $1 \cdot 10^5$ particles into the flow
and integrated the system for approximately 4 large eddy turnover
times. The statistics are computed from the last 3 large eddy turnover
times for all runs.
\section{Eulerian statistics}
\begin{figure}
  \centering
  \includegraphics[width=0.8\textwidth]{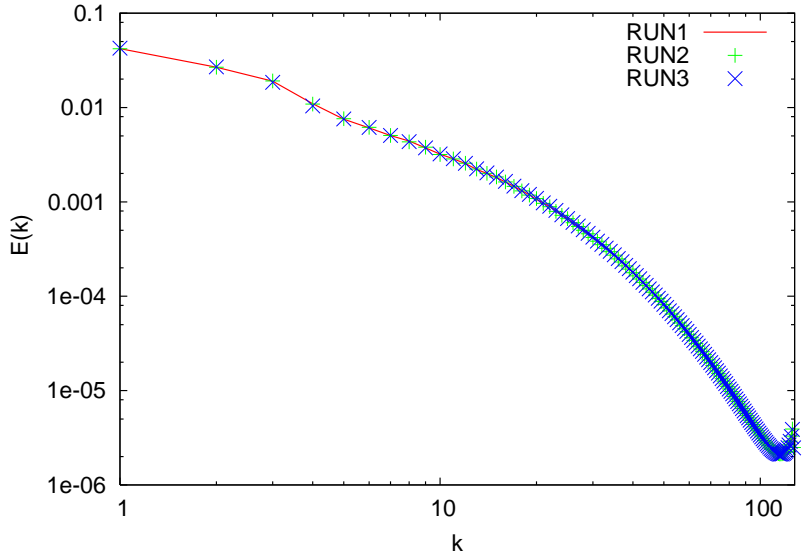}
  \caption{Energy spectra from RUN1 -- RUN3}
  \label{energySpec}
\end{figure}
\begin{figure}
  \centering
  \includegraphics[width=0.8\textwidth]{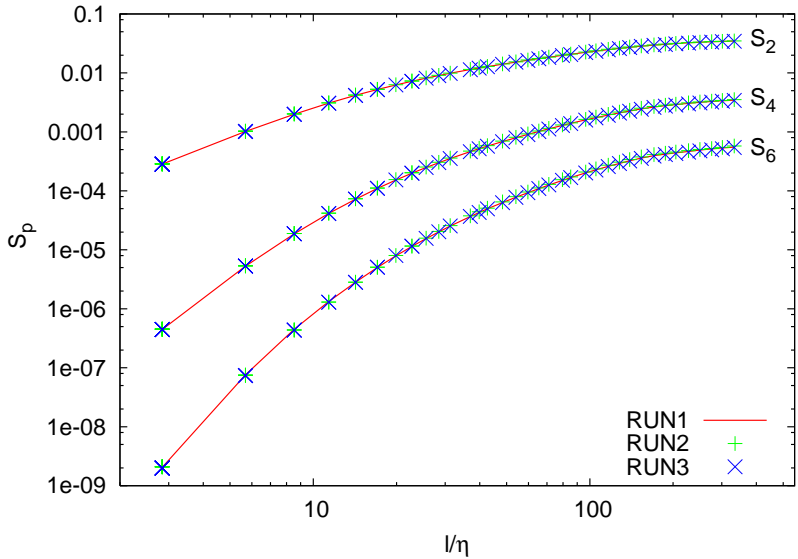}
  \caption{Eulerian longitudinal structure functions $S_2$, $S_4$ and
  $S_6$ (see (\ref{eulerStrucFunc})) from top to bottom from RUN1 --
  RUN3}
  \label{eulerStruc}
\end{figure}
\begin{figure}
  \centering
  \includegraphics[width=0.8\textwidth]{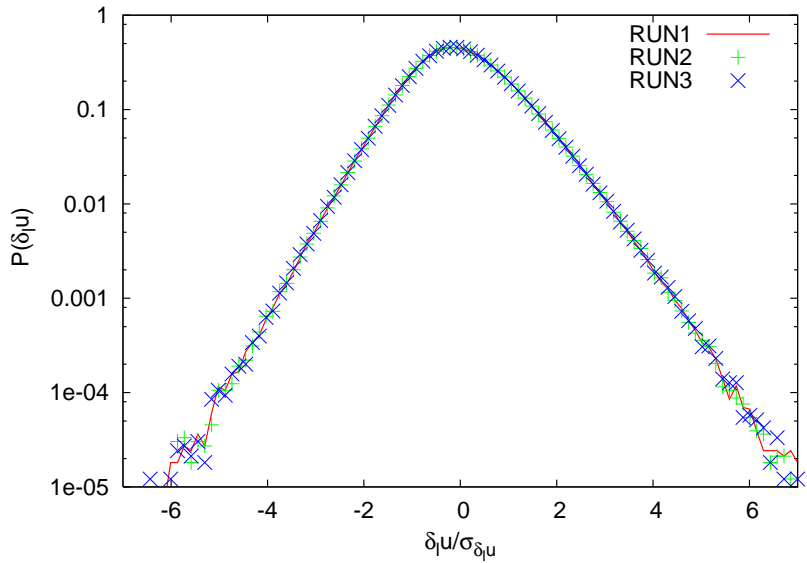}
  \caption{Eulerian probability density function for
    $u_x(x+2dx)-u_x(x)$ from RUN1 -- RUN3, normalized to unit
    variance}
  \label{eulerPDF}
\end{figure}

A central point of many theories of turbulence \cite{kolmogorov:1991}
are the scaling laws of the energy spectrum within the inertial
range. Figure~\ref{energySpec} shows the computed energy spectra from
RUN1 -- RUN3. Because of the limited Reynolds number, no clear scaling
range is visible. More important for the purpose of this paper is the
fact that the spectra of the different runs are hardly
distinguishable. Even higher order statistics such as the Eulerian
longitudinal structure functions
\begin{equation}
  \label{eulerStrucFunc}
  S_p(l) =
  \left<|(\u(\x+\mathbf{l})-\u(\x)) \cdot \mathbf{\hat{l}}|^p\right>,
\end{equation}
angular brackets denoting spatial averaging, look the same for all
types of floating-point precisions (see Figure~\ref{eulerStruc}).
A more subtle comparison yields the according Eulerian probability
density function (PDF) of velocity
increments. Figure~\ref{eulerPDF} shows the PDFs for the variable
$u_x(x+2dx)-u_x(x)$ normalized to unit variance.
The PDFs differ just within the statistical errors due to the finite
statistical ensemble, but the overall shape is identical. Therefore
the floating-point precision has no impact on the considered Eulerian
statistical properties of turbulent flows. One can suspect that also
other Eulerian statistical quantities are unaffected by the underlying
floating-point precision.

\begin{figure}
  \centering
  \includegraphics[width=0.8\textwidth]{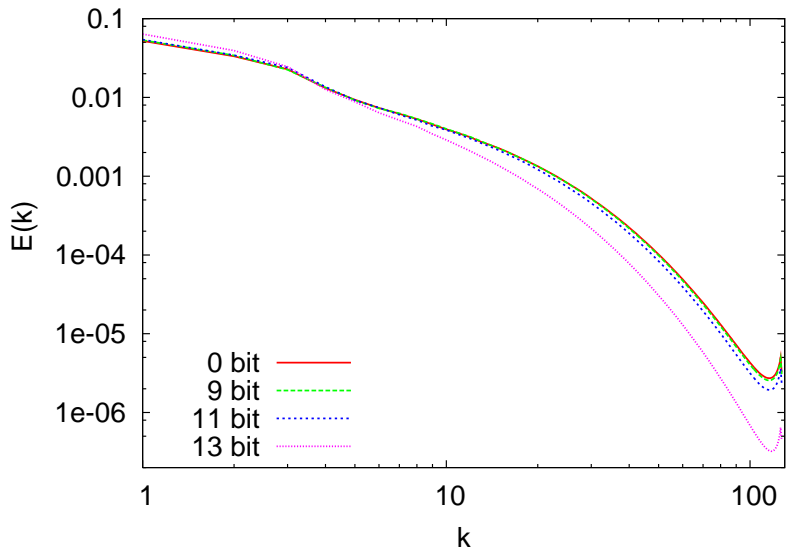}
  \caption{Energy spectra for simulations with single floating-point
  precision reduced by several bits}
  \label{energySpecBit}
\end{figure}

\begin{figure}
  \centering
  \includegraphics[width=0.8\textwidth]{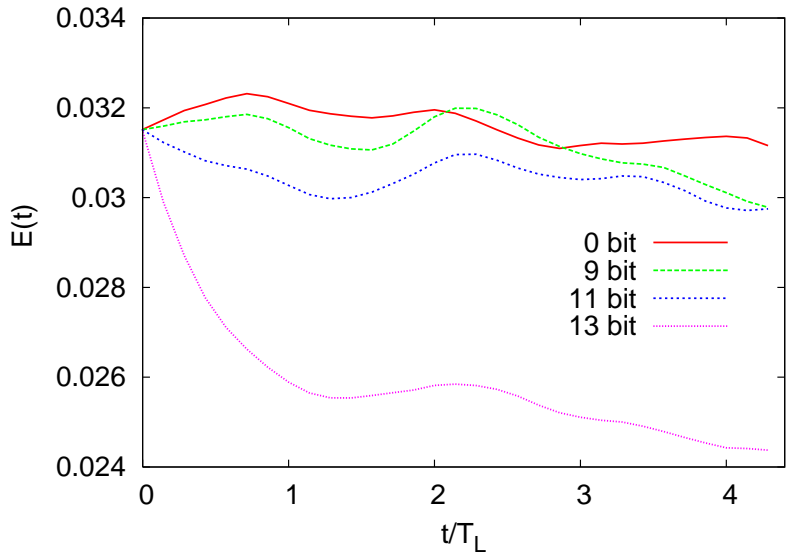}
  \caption{Fluctuating total energy for simulations with single
  floating-point precision reduced by several bits}
  \label{energyBit}
\end{figure}

Now the question will be addressed up to which grid resolution these
results remain true. A rough estimate can be given by determining the
threshold at which the statistics are affected by the floating-point
precision. Starting from single floating-point precision for all
fields we artificially reduced the precision by cutting off several
least significant bits of the mantissa. Figure~\ref{energySpecBit}
shows energy spectra from simulations with single and reduced
floating-point precision. Reducing the floating-point precision by
nine bits yields unaffected statistical results, while a reduction of
11 bits slighty spoils the spectrum. Cutting off 13 bits yields a
clearly modified energy spectrum. These findings are similar for
high-order structure functions. The threshold is therefore
approximately nine bits which can also be seen from
Figure~\ref{energyBit}. Here the temporal evolution of the total
energy is depicted. Besides the fluctuations, the dynamics are
affected at a reduction of 11 bits.

In order to give an estimate up to which grid size high Reynolds number
simulations can be performed in single precision, where the ratio of
the grid spacing $dx$ to the dissipation length scale $\eta$ is kept
constant, we assume a Kolmogorov scaling of the energy spectrum. This
means that the energy range will be enlarged by a factor of $n^{5/3}$
when increasing the grid resolution by a factor of $n$ in each
direction. Enlarging a number by nine bits yields a 512 times larger
number. We therefore expect that it is still possible to use single
floating-point precision data with $4096^3$ grid points.

\section{Lagrangian statistics}
\begin{figure}
  \centering
  \includegraphics[width=0.8\textwidth]{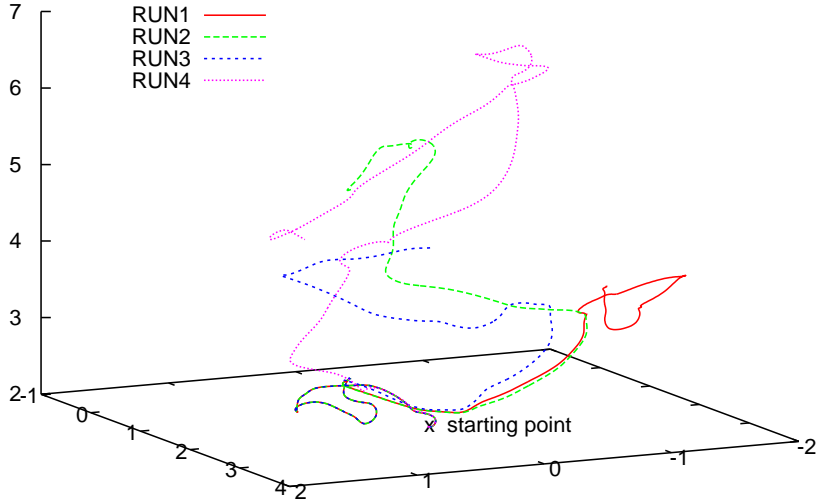}
  \caption{Trajectories of a single particle from RUN1 -- RUN4}
  \label{fig_traj}
\end{figure}

Apart from the Eulerian formulation where the evolution of the
velocity field is considered at fixed points in space, the Lagrangian
coordinates follow the fluid elements in time. Figure~\ref{fig_traj}
shows trajectories of a single particle starting at x for RUN1 --
RUN4.
The particles are seeded into the flow at the same time and same
positions for all runs and are followed for approximately 4 large-eddy
turnover times. The trajectories differ substantially. RUN2 stays
close to RUN1 for the longest time as one would expect. The sudden
aberration of RUN2, RUN3 and RUN4 from RUN1 is due to the chaotic
character of the turbulent flow. The trajectory of RUN4 deviates first
from the others. This already implies that the interpolation scheme has a
greater impact on the trajectories than the floating-point precision.

\begin{figure}
  \centering
  \includegraphics[width=0.8\textwidth]{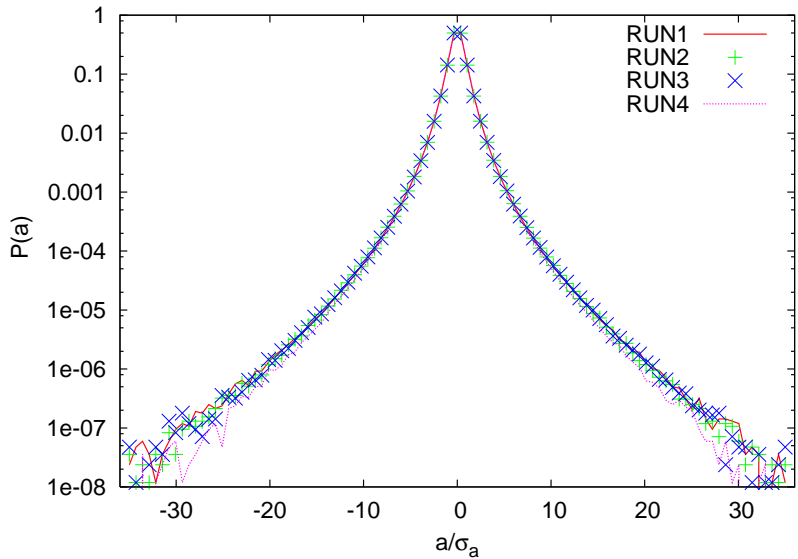}
  \caption{PDFs of Lagrangian velocity increments (\ref{lagIncrement})
    from RUN1 -- RUN4, normalized to unit variance}
  \label{fig_lag_pdf}
\end{figure}

\begin{figure}
  \centering
  \includegraphics[width=0.8\textwidth]{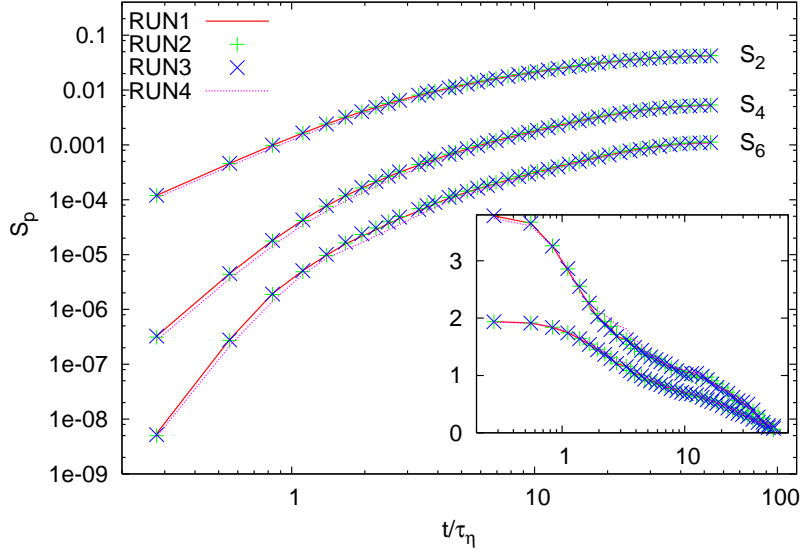}
  \caption{Lagrangian structure functions (\ref{lagStrucFunc}) from
  RUN1 -- RUN4, inset: logarithmic derivative of $S_2$ (bottom) and
  $S_4$ (top)}
  \label{fig_lag_struc}
\end{figure}

\begin{figure}
  \centering
  \includegraphics[width=0.8\textwidth]{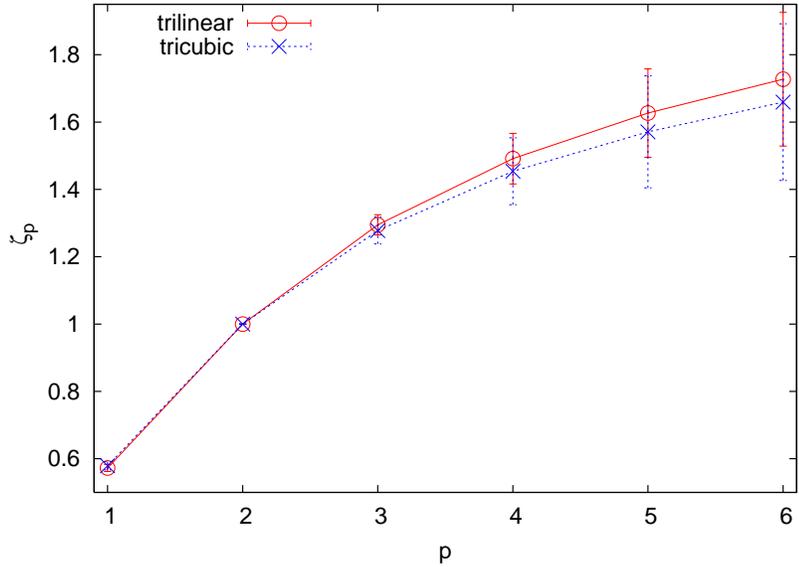}
  \caption{Lagrangian scaling exponents using ESS for the tri-linear
  (RUN4) and tri-cubic (RUN1) interpolation scheme}
  \label{ess}
\end{figure}

To analyze the impact of differing single particle trajectories on the
statistics of an ensemble of fluid elements we computed the PDFs of
the Lagrangian acceleration (acceleration of fluid elements), which
equals the PDF of the Lagrangian velocity increments
\begin{equation}
  \label{lagIncrement}
  |(\u(t+\tau)-\u(t))\cdot \hat{\e}|
\end{equation}
for small time increments $\tau$.  The velocity increment after a
time-lag $\tau$ is projected onto a coordinate axis $\hat{\e}$. These
PDFs for RUN1 -- RUN4 are shown in Figure~\ref{fig_lag_pdf}.
The differences between the runs with different floating-point
precisions differ just within the statistical fluctuations. The PDF
computed with the tri-linear interpolation in RUN4 is slightly more
narrow than the PDF with the tri-cubic interpolation in RUN1. This is
because the tri-cubic interpolation scheme is more capable to follow
the trajectories of the nearly singular structures (vortex tubes)
which are presumably responsible for the stretched tails of the PDFs
\cite{mordant:2004}. As the broadness reflects the degree of
intermittency and the Reynolds number of the turbulent flow
\cite{vincent-meneguzzi:1991}, the tri-linear interpolation scheme
might underestimate the degree of
intermittency. Figure~\ref{fig_lag_struc} shows the corresponding
Lagrangian structure functions
\begin{equation}
  \label{lagStrucFunc}
  S_p(\tau)=\left<|(\u(t+\tau)-\u(t))\cdot \hat{\e}|^p\right>,
\end{equation}
angular brackets denoting temporal averaging. These functions clearly
exhibit no differences for RUN1 -- RUN3, i.e. depending on the
floating-point precision. Concerning the interpolation scheme, the
Lagrangian structure functions slightly differ for RUN1 and RUN4. As
for the PDFs, the interpolation scheme has a small impact on the shape
of the measured Lagrangian structure functions.  The inset of
\ref{fig_lag_struc} shows the logarithmic derivative of the second and
fourth order structure function. Due to the limited Reynolds number no
clear scaling region is observable and no absolute scaling exponents
can be extracted. We computed the relative scaling exponents using the
assumption of extended self-similarity (ESS) (see
\cite{benzi-ciliberto-etal:1993b}). Figure~\ref{ess} shows the
relative scaling exponents. Despite the large error-bars the linear
interpolation scheme systematically yields less intermittent
statistics than the cubic interpolation scheme. This is in agreement
with the observation concerning the Lagrangian PDFs. The tri-cubic
interpolation scheme results in more stretched tails than the
tri-linear scheme. An increased degree of intermittency is reflected
in a more intermittent scaling behavior of the Lagrangian structure
functions. The higher the order of the scaling exponent the larger is
the difference between the interpolation schemes. High-order structure
functions are determined by the most intense events resulting from the
most singular structures in the flow. These are spiral motions close
to strong vortex filaments. Our findings show that the interpolation
scheme has an influence on the computation of the Lagrangian scaling
behavior.
It has to be stressed that these findings also remain true for
simulations using less or more collocation points if the ratio of the
dissipation length scale and the grid spacing is similar to that of
the simulations presented. This ratio determines the smoothness of
discretized velocity field. For higher ratios the differences due to
the interpolation scheme might become smaller. However, high Reynolds
number simulations commonly use ratios similar to the one used in this
work. Thus, the accuracy of the interpolation scheme will have an influence
on the statistical results independently of the number of collocation
points.

\section{Conclusions}
We investigated the impact of the floating-point precision and the
interpolation scheme on the results of DNS of turbulence by
pseudo-spectral codes. To analyze the influence of the floating-point
precision, we performed three runs. First we used double precision for
all fields (RUN1). Second, we used single precision for the velocity
fields and double precision for the convolution fields (RUN2), like
the currently largest simulation of turbulence on the Earth Simulator
did. The third simulation uses single precision for all fields (RUN3),
which halves the amount of memory needed compared to RUN1 and
therefore allows for an increased Reynolds number. Although the
trajectories of tracer particles differ depending on the
floating-point precision which is due to the chaotic nature of
turbulence, the statistical quantities such as the energy spectrum,
the Eulerian and Lagrangian structure functions and the corresponding
PDFs show only minor, negligible differences. Therefore it is possible
to achieve the same statistical results by halving the amount of
memory needed.

The differences according to the interpolation scheme are more
pronounced. Again, the single particle trajectories differ
substantially between the tri-cubic and the tri-linear
scheme. Furthermore the Lagrangian PDFs have slightly differing
shapes. The tri-cubic interpolation scheme reflects a higher degree of
intermittency than the tri-linear one. This is because the first is
more capable the resolve the nearly singular structures (vortex
tubes). These are responsible for the intermittency of the flow. The
Lagrangian structure functions and their logarithmic derivative show
slightly different shapes and scaling behavior. The tri-cubic
interpolation scheme yields a slightly more intermittent statistic
than the tri-linear scheme.

\ack Access to the JUMP multiprocessor computer at the FZ J\"ulich was
made available through project HB022. Part of the computations were
performed on an Linux-Opteron cluster supported by HBFG-108-291. This
work benefitted from support through SFB 591 of the Deutsche
Forschungsgesellschaft.

\bibliographystyle{elsart-num}
%\bibliography{master}

\end{document}